\documentclass[10pt]{article}

\usepackage[fleqn]{amsmath}   
\usepackage{amssymb}
\usepackage{graphicx}
\usepackage{cite}
\usepackage{color} 

\usepackage{setspace} 
\usepackage[labelfont=bf,labelsep=period,justification=raggedright]{caption}

\makeatletter
\renewcommand{\@biblabel}[1]{\quad#1.}
\makeatother

\date{}

\usepackage{tabularx}  % automatically adjusts column width in tables
\usepackage{multirow}  % allows entries spanning several rows  
\usepackage{colortbl}  % allows coloring tables

\newcommand{\hdr}[3]{%
  \multicolumn{#1}{|l|}{%
    \color{white}\cellcolor[gray]{0.0}%
    \textbf{\makebox[0pt]{#2}\hspace{0.5\linewidth}\makebox[0pt][c]{#3}}%
  }%
}

\begin{document}

\begin{flushleft}
{\Large
\textbf{A Behavioural Perspective on the Early Evolution of Nervous Systems: A Computational Model of Excitable Myoepithelia}
}
\\
Ronald A. J. van Elburg$^{1,\ast,\dagger}$, 
Oltman O. de Wiljes$^{1,2,\dagger}$, 
Michael Biehl$^{3}$,
Fred A. Keijzer$^{2}$
\\
{\bf{1}} Institute of Artificial Intelligence, Faculty of Mathematics and Natural Sciences, University of Groningen, Groningen, The Netherlands
\\
{\bf{2}} Department of Theoretical Philosophy, Faculty of  Philosophy, University of Groningen, Groningen, The Netherlands
\\
{\bf{3}} Johann Bernoulli Institute for Mathematics and Computer Science, Faculty of Mathematics and Natural Sciences, University of Groningen, Groningen, The Netherlands
\\
$\ast$ E-mail: RonaldAJ@vanElburg.eu
\\
$\dagger$ These authors contributed equally.
\end{flushleft}

\section*{Abstract}
How and why the first nervous systems evolved remain open questions. One influential scenario casts excitable myoepithelia (epithelia that combine conductive and contractile properties) as a plausible proto-nervous system. We argue that while modern myoepithelia rely on gap junctions, early myoepithelia had to rely on paracrine signalling or equivalently local chemical transmission and constitute a crucial step towards modern nervous systems. Our main questions concern the coordinative possibilities and limitations of such excitable myoepithelia and their potential relevance as an intermediate step to nervous systems.

We used conductance based model cells to create artificial myoepithelia with various shapes and electrophysiological characteristics. We developed a measure for whole body coordination in the myoepithelium activity pattern. Using this measure we show that excitable myoepithelia relying on paracrine signalling can exhibit body-scale patterns of activation. Relevant factors determining the extent of patterning are the noise level for spontaneous vesicle release, relative body dimensions, and body size. In small myoepithelial models whole-body coordination emerges from cellular excitability and excitatory chemical transmission alone. At larger body sizes the intrinsic noise of chemical transmission limits whole-body coordination.

We speculate that while proto-neural myoepithelia could have provided a solution for basic forms of muscle-based movement, there would have been a strong evolutionary pressure to improve on this mechanism by (a) the development of non-chemical transmission mechanisms and (b) a switch to nervous systems proper by including axodendritic processes. Our study supports a two-step evolutionary scenario for nervous systems. In the first step chemical receptors and the unicellular machinery for exocytosis evolve into paracrine signalling, thus providing multicellulars with whole-body coordination. In the second step, axodendritic processes evolve under the evolutionary pressure towards larger body sizes.

\section*{Introduction}
The evolution of the first nervous systems raises fundamental questions. Nevertheless, it has received comparatively little scientific attention and there is no consensus as to how nervous systems first evolved \cite{miller_2009, moroz2009independent, lichtneckert2007origin}. One reason for this situation has been the dearth of relevant scientific data that can be made to bear on these early events. Nowadays, molecular and genetic techniques have changed the situation at a molecular level, although less progress has been made concerning the macroscopic features of early nervous systems. In the present study, we introduce a computational approach to investigate the potential behavioural relevance of an early (proto-)neural organization: excitable myoepithelia.

\subsection*{Evolution of the earliest nervous systems}
Modern molecular genetic approaches to evolutionary biology yield time estimates for the first appearance of the different molecular building blocks of modern neurons. It is now increasingly clear that key molecular components like ion channels, neurotransmitters and synaptic protein families must have been present in precursor organisms without nervous systems and even in single-celled organisms \cite{greenspan2007introduction, Ryan_Grant_2009}. It is also clear that the outlines of the genetic signalling devices involved in nervous system development have been present from a very early start, some components also being present in single-celled precursors \cite{arendt_2008}, while genomic studies have uncovered that Cnidaria (e.g. jellyfish), which is the earliest divergent phylum to have a nervous system, possess an almost complete set of signalling molecules that have critical roles in bilaterian neurodevelopment \cite{Watanabe_Holstein_2009a}. 

However, progress at the level of animal behaviour and nervous system anatomy has been much slower and proposed ideas remain inconclusive \cite{miller_2009,mackie1990elementary,lichtneckert2007origin}. Work here so far consists mostly of different evolutionary scenarios explicating how excitable cells developed processes and synapses, became connected and evolved into basic forms of nervous systems that still exist today. Lichtneckert and Reichert~\cite{lichtneckert2007origin} discuss the main proposals for the evolution of nerve cells and nervous systems and, like Mackie\cite{mackie1990elementary} before them, conclude that it seems impossible to rate any of these proposals as more plausible or relevant than the others. 

The problem is that on the basis of the fossil record alone such proposals must remain speculative and cannot be tested or validated in any conclusive way. It is unknown whether the organisms involved looked or behaved like any modern animal. Even the most `primitive' nervous systems in existence today ---those of cnidaria--- are highly evolved \cite{mackie_2004} and cannot be directly taken as a model for the morphological and organizational properties of the organisms that first evolved nervous systems. According to conservative molecular estimates, the event of the evolution of the first nervous system ---as indicated by the first divergence wthin the eumetazoa--- happened more than 600 million years ago \cite{peterson2008ediacaran}, while other molecular estimates push the event back to a 1000 million years ago \cite{blair2009animals}. In either case, we do not have solid ---or indeed any--- fossil evidence concerning morphology, life style or behaviour \cite{Valentine_2007,erwin2006}, except for the fact that the animals involved must have been below the millimetre range \cite{Brasier_2009}. It is even possible that nerve cells evolved several times independently \cite{moroz2009independent}, involving different organisms. Given these uncertainties, we think that developing additional sources of evidence concerning this crucial initial evolutionary step should receive a high priority. 

In this paper we add computational modelling to the existing techniques of gathering evidence on the early evolution of nervous systems. Computational modelling allows for a systematic investigation of the possible macroscopic operation of the early organisms that evolved nervous systems. Most notably, computational models enable us to investigate how electrophysiological, cell-signalling and biomechanical properties of simple multicellular organisms can self-organize into whole-body behaviour and coordination. While these models do not directly shed light on what actually happened, they allow us to say more about the possibilities available for organization and coordination during early nervous system evolution. Additionally, these models help clarify the relevance of nervous systems and their precursors for the behaviour of simple multicellular organisms. In particular, the identification of behavioural options and limitations of evolutionary precursors to the nervous system might help answer the question: What drove the early evolution of the nervous system?

\subsection*{Excitable myoepithelia as potential proto-nervous systems}
As regards the possible role of early nervous systems in organizing behaviour, the first question to ask is in what context early nervous systems may have developed. While the  most notable feature of modern nervous systems is their involvement in sensory information processing and subsequent action execution \cite{Watanabe_Holstein_2009a,Jekely2011}, the setup and operation of early nervous systems and any immediate precursor systems may have been different. In the present paper we build on the general finding that the evolution of nervous systems is intertwined with the evolution of muscle as the primary source of motility in eumetazoa \cite{Seipel_Schmid_2005}. Muscles and nervous systems come together in evolution and it seems plausible that nervous systems are deeply implicated in the evolutionarily important switch from movement based on cilia to muscle-based movement.

A highly relevant idea in this context comes from Pantin~\cite{pantin1956origin}, who argued that the operation of the first nervous system was to organize and coordinate body movement based on muscle contractions giving rise to what he called the `metazoan behaviour machine'. Pantin claimed that this new effector, consisting of extended surfaces that had to contract in a coordinated way to enable movement, required a mechanism of large-scale coordination, providing a clear proximate reason for the evolution of nervous systems as such coordinators. Passano~\cite{passano1963primitive} added that such a neural organization also required endogenous pacemakers to keep it going. Direct sensorimotor connections, such as the reflex arc, were only a later development in this view.

However, the later discovery of excitable epithelia \cite{Mackie1965} seemed to undercut this explanation. The epithelium of an animal can act as an excitable sheet where action potentials, once initiated, travel in all directions. In modern cases, these epithelial cells are linked by direct cytoplasmic connections or gap junctions \cite{anderson1980epithelial, Josephson_1985}. Seemingly, whole-body coordination of extensive muscle sheets can be accomplished by an excitable epithelium closely linked to a sheet of contractile tissue, or even by a single excitable myoepithelium that combines excitable and contractile (myoid) functions. For example, in the hydrozoan species \emph{Sarsia} and \emph{Euphysa}, the `subumbrellar ectoderm is a single layer of cells having striated fibres running circularly' forming `the swimming muscle'~\cite{mackie1968epithelial}, in this epithelial tissue electrical impulses travel from cell to cell via gap junctions and are able to induce effector responses at points distant from the signals' origin. Similar forms of excitable tissue are also present in muscle~\cite{Josephson_1985,mackie2004epithelial,Brink1996}, plant fibres \cite{masi2009spatiotemporal}, sponge syncytial tissues \cite{leys1999impulse} and myocardial tissues~\cite{nash2004electromechanical, tusscher2006alternans, tusscher2007organization}. For basic muscle coordination, it seems, there is no obvious need for a nervous system. 

Nevertheless, the  claim that large scale muscle coordination could evolve without any form of nervous system can be challenged. Two issues are important. What is a nervous system in the first place? Second, are modern excitable epithelia a primitive condition or a later development? We will discuss these two points in turn.

How do we define nervous systems? Current views take neurons as a package deal: `a typical neuron has four morphologically defined regions: cell body, dendrites, axon and presynaptic terminals' \cite{KSJ_Essentials}. However, it is plausible to separate the evolution of chemical transmission from the evolution of long distance processes~\cite{mackie1990elementary,lichtneckert2007origin}. In this case the evolution of the first nervous systems consists of at least two separate evolutionary steps:
\begin{enumerate}
\item[i.] The evolution of chemical transmission, allowing cells to pass electrical signals to adjacent cells.
\item[ii.] The evolution of axodendritic processes that enable electrical signals to be sent to nonneighbouring cells.
\end{enumerate}
Subsequently, the evolution of the first (proto-)nervous systems can be taken as the evolution of the first excitable epithelia based on electrical signalling through chemical transmission but, so far, without long-distance axodendritic processes. From these evolutionary considerations one can draw the conclusion that an excitable system based on chemical transmission but without processes is already a proto-nervous system. Thus, the origins of the first nervous systems can be traced to the evolution of such excitable epithelia themselves.

Current molecular studies suggest that the molecular machinery for chemical transmission predates multicellulars. In fact already in early unicellular animals genes coding for glutamate receptors and intracellular calcium signalling were found~\cite{Ryan_Grant_2009}. In addition, the SNARE protein, an important molecule for transmembrane transport, was available in unicellulars as well~\cite{Kloepper_2008}. Furthermore, voltage-gated ion channels, crucial for excitability, are thought to be derived from a prokaryotic ancestor~\cite{Anderson_2001,Derst_1998} and thus are much older than multicellular animals and therefore older than neural systems. 

In contrast, the molecular basis of gap junctions (connexins and pannexins) seems to have arrived after multicellularity, and even today starlet sea anemones (\emph{Nematostella vectensis}) have no genes coding for these proteins~\cite{Shestopalov_2008}. Additionally, electrophysiology could not establish gap junctions in certain other cnidarian species like anthozoans (sea anemones like \emph{Aiptasia pulchella}) and scyphozoans (jellyfish like \emph{Cyanea capillata}) although similar methods revealed gap junctions in hydrozoans~\cite{Mackie_1984,Satterlie2011}. 

In view of the above we postulate that excitable (myo)epithelia relying on local chemical transmission may very well have been the first major step towards modern nervous systems. Hence, to clarify the possible operation of such a proto-neural setup we need to know how excitation spreads through local chemical transmission. In this paper, we describe a computational model of such an excitable myopepithelium aimed at describing this spread of excitation. Our model is comparatively abstract at the macroscopic scale (making as few assumptions as possible concerning the animal involved) while incorporating as much of the biomolecular and physiological details as necessary. In this way, we can study the spreading of activity on such an excitable myoepithelium which acts both as a signalling device and a primitive muscle-based effector. In  particular, we aimed to understand the influence that animal size and shape exert on whole-body coordination.

To our knowledge, no earlier modelling studies addressing the behaviour of excitable myoepithelia exist. A somewhat similar model was used to describe travelling waves on coral nerve networks that extend across several polyps~\cite{Chen_TravWavesCoral_2008}. However, that study targeted modern nerve nets rather than the basic myoepithelium of the present study. On a more abstract level our model is related to experimental and theoretical work on Mexican waves and related phenomena~\cite{farkas2002social}, which highlights the self-organized patterns of activation across an excitable medium. Our study provides one of the first computational models of early nervous system functioning and evolution. In this respect, we regard this study as a groundbreaking application of computational neuroscience to evolutionary neurobiology.
\label{sec:intro}

% You may title this section "Methods" or "Models". 
% "Models" is not a valid title for PLoS ONE authors. However, PLoS ONE
% authors may use "Analysis" 
\section*{Methods}
We developed our model to establish under which conditions a surface of excitable cells can generate coordinated patterns of activity suitable for the control of body movement through muscle contractions. In myoepithelia coordinated electrical activity across the epithelium necessarily involves coordinated contractions, we modeled the initiation and spreading of these electrical excitations. To analyse the model outcomes we developed two indicators of whole-body coordination. These indicators can summarize simulation outcomes in a form suitable for visualizing the outcomes of parameter scans over  the circumference and the length of the networks to be introduced shortly.   

\subsection*{Model}

\begin{figure}
\begin{center}
\includegraphics[width=0.5\textwidth]{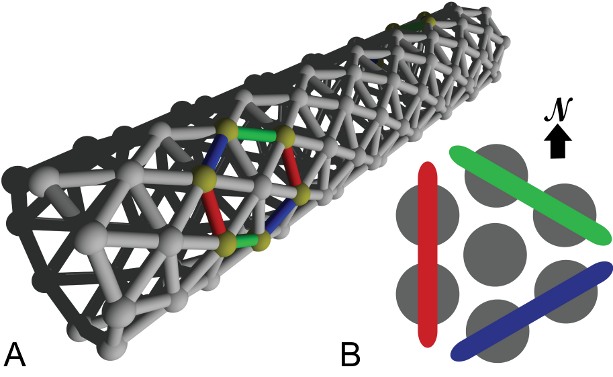}
\end{center}
\caption{Model Network A. Three dimensional organization of our model network, resembling a tube-shaped animal. The myoepithelium model consists of excitable cells arranged into a triangular lattice wrapped around a cylinder. B. Colour coding used to refer to the three differently oriented wave fronts on this lattice. The size of a wave front is established by counting the number of neighbouring cell pairs on the corresponding wave front.} 
\label{fig:ModelNetwork}
\end{figure}

\label{sec:mm}

Our cell network model is built from single-compartmental cell models endowed with standard Hodgkin-Huxley dynamics~\cite{HODGKIN1952a}; for parameter details see our Nordlie-Gewaltig-Presser-style \cite{Nordlie2009} model summary tables~\ref{table:MS-A} to ~\ref{table:MS-H}. These model cells were arranged in a triangular lattice generated by a purpose-built weight generator. This generator was designed for a network building library (Available upon request from the corresponding author) developed by one of us for the NEURON simulation environment~\cite{carnevale_hines_2006}. To facilitate parameter scanning, the multiple run control~\cite{elburg2010impact} made available in ModelDB (http://senselab.med.yale.edu/ModelDB, accession number 114359) was used. 
The network consists of a sheet of cells on a triangular lattice rolled up to a cylinder. We decided to use a triangular lattice because it is more isotropic than other two-dimensional lattices and therefore introduces the least amount of directional bias. Excitatory chemical transmission between nearest neighbours including the dynamics of messenger molecules exocytosis and receptor channel kinetics are modelled as double exponential conductance changes in receptor channel populations~\cite{destexhe1994synthesis}. Only a single type of excitatory chemical transmission is included. 
An important aspect of this model is the rate of spontaneous vesicle release. Data on spontaneous release frequencies of vesicles are still scarce and we found only one such study~\cite{Mackenzie2000}. Experimentally, the maximum spontaneous vesicle release rate found was $0.25$ Hz. In our model, which has integrated six chemical transmission sites into a single model synapse, this yields a total spontaneous vesicle release rate of $1.5$ Hz per model synapse. A reliable lower bound, other than zero, on the vesicle release rate is not available because many synapses failed to show spontaneous vesicle release during the experiment. From the biophysics of vesicle release we further know that vesicle release is calcium-concentration dependent~\cite{Augustine2001}. As calcium concentration dynamics is known to vary with surface-to-volume ratio and with the concentration, mobility and kinetics of the endogenous calcium binding proteins~\cite{Cornelisse2007}, we can expect a large range of vesicle release rates. We have therefore chosen to vary model vesicle release rates over 5 orders of magnitude around a value of 0.1 Hz. Thus we included the maximum directly observed vesicle release frequency in our range, but have a strong bias towards lower vesicle release rates. 

We also investigated the possibility to introduce spontaneous network activity through spontaneous spiking resulting from stochastic ion channel gating. However, a short exploration using the model developed by \cite{Linaro2011} (available from ModelDB accession number: 127992) showed that this would lead to spike rates much lower than those induced with the vesicle release rates included in the model. With this in mind and in view of the computational cost we refrained from including ion channel gating in our model.

\subsection*{Analysis}

\begin{figure}
\begin{center}
\includegraphics[width=0.5\textwidth]{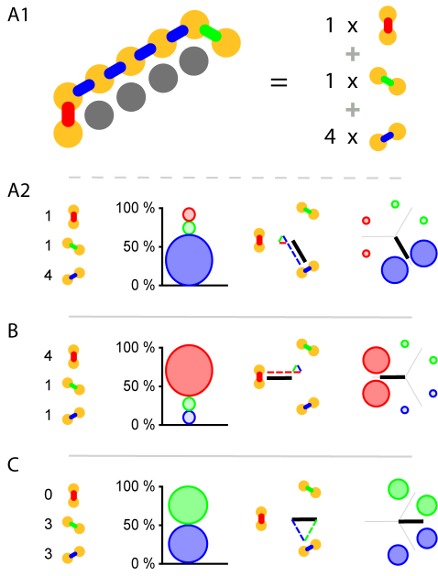}
\end{center}
\caption{Visual representations used. (A1) Yellow disks represent active cells at a given time; grey cells represent inactive cells. Neighbouring pairs of active cells come in three different orientations and are arbitrarily labelled with red, green and blue. To establish which orientation is dominant we simply count the occurrence of the orientation labels. (A2) The orientation label counts (left) are translated into relative wave-front orientation prevalences (graph left of middle) and represented by the  diameters of the disks with the corresponding colour labelling. In addition these counts are translated into an average wave-front propagation orientation by adding the normal vectors to these wave fronts with a weight proportional to their label count (diagram right of middle). Relative wave-front orientation prevalences and propagation direction are combined into a single representation for use in parameter scans (right). (B,C) Like A2 with different orientation label counts.} 
\label{fig:wavefrontanalysis}
\end{figure}

As whole-body coordination is not a well-defined mathematical concept at present, it is crucial that we should choose good indicators of it. The triangular lattice supports three possible wave-front orientations; as indicators we have chosen the relative amounts with which these orientations appear in our simulations. Subpanel A1 of figure~\ref{fig:wavefrontanalysis} illustrates our analysis method and shows that pairs of neighbouring cells come in three different orientations: North--South, North East--South West, and South East--North West. For each of these orientations we count the number of neighbouring pairs that fire within 2 ms of each other. In subpanel A2 of the same figure we show how these raw counts (left) are translated into percentages (middle left), which are then used to set the diameter of the circles in the oriented circle pairs (middle left and right). This presentation, which we call \textbf{relative wave-front orientation prevalence}, is suited for the analysis of parameter scans, e.g. figures~\ref{fig:WavefrontPrefsNoise0_1} and~\ref{fig:WaveOrientation_NoiseRangeExtremes}. We used this representation to present averages over all runs at a specific parameter setting. 

Wave fronts propagate roughly perpendicularly to their own orientation. This idea leads to a second representation. Instead of showing the percentages directly, we add up the vectors normal to the wave fronts weighed by the same percentages used in our relative wave-front orientation prevalence representation. For the North--South oriented wave-front, propagation is to the East or the West, i.e. parallel to the normal vector pointing West, similarly propagation is parallel to a South East-pointing vector for the North East--South West-oriented wave front, and parallel to North East pointing vector for the South East--North West-oriented wave front. The choice of these normal vectors is not unique, we selected them in such a way that they point from the center to the oriented pair in the relative wave-front orientation prevalence representation. In figure~\ref{fig:wavefrontanalysis}:A2 middle right we show the vector addition and the resulting vector. We call this representation \textbf{wave-front propagation orientation}. This representation is also suitable for the presentation of parameter scans and additionally allows us to show both the individual simulation runs and the average over simulation runs in a single figure. Supplementary figure~\ref{fig:WavefrontIndividualRuns} uses this representation to show that we obtain similar results over twenty runs in which only the random number generator initialization is changed. 

For the purpose of this study, visual inspection of our earlier simulations showed that both relative wave-front orientation prevalence and average wave-front propagation orientation are reasonably good indicators of the effects of body size and chemical transmission noise on whole-body coordination.  

% Results and Discussion can be combined.
\section*{Results}

\begin{figure}
\begin{center}
\includegraphics[width=0.8\textwidth]{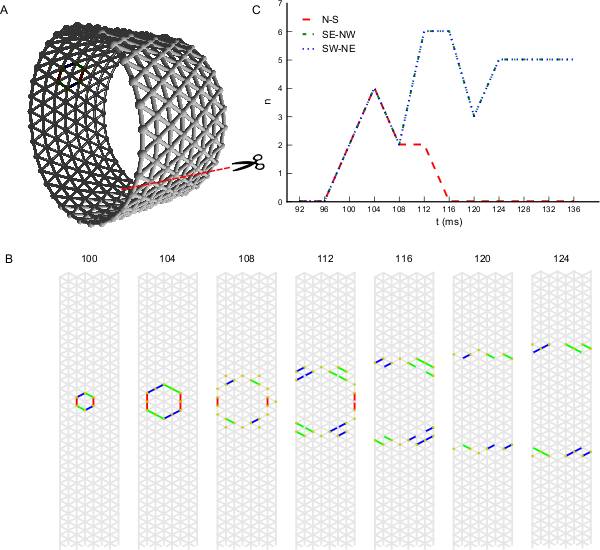}
\end{center}
\caption{Wave patterns on a short tube (length: 8 cells, circumference: 32 cells, noise rate: 0.1 Hz). (A) Network geometry: the scissors indicate the line at which the tube is cut for presentation in B. (B) Snapshots of network activity during 4 ms intervals in an illustrative phase of the dynamics. The wave fronts propagating longitudinally die out at the network edges, causing the North--South oriented wave fronts to disappear. As a result, wave-fronts propagate predominantly transversely to the tube and the wave-front orientations are South East--North West and South West--North East. (C) Temporal development of different wave-front orientations, including all  snapshot times shown in B.} 
\label{fig:shortpipe}
\end{figure}

In subpanel B of figure~\ref{fig:shortpipe} we see the temporal development of wave-front patterns on the short cylindrical network also shown three dimensionally in subpanel A of the same figure. After initial excitation the wave fronts grow in size uniformly in all directions until the wave fronts propagating longitudinally reach the edge of the cylinder and disappear. The remaining wave fronts propagate in both transverse directions. Provided no other noise-induced wave fronts interfere, these wave fronts eventually annihilate each other on the side opposite the wave front initiation point. In subpanel C of figure~\ref{fig:shortpipe}, where the wave-front orientation counts are shown for the time interval depicted in subpanel B, we clearly see the phenomena we just described reflected. Initially all wave-front counts grow at the same rate, then the North--South oriented wave front, propagating longitudinally, dies out and the remaining wave fronts continue to grow in size. Wave-front counts are approximately stable during transverse propagation. The double wave fronts visible at later times occur because the data in this figure are binned in $4$ ms bins, while a pair of neighbouring cells is considered to be part of a wave front if they fire within $2$ ms of each other. Also visible is the breaking up of an initial wave front into smaller subfronts. We consistently observe slower spreading of the wave fronts at points where the wave front changes orientation and the two differently oriented parts move apart. Similarly we see a faster spreading of the wave fronts at points where the wave front changes orientation and the two differently oriented parts move towards each other. This dynamics tends to flatten wave fronts.

\begin{figure}
%\begin{center}
\includegraphics[height=16cm]{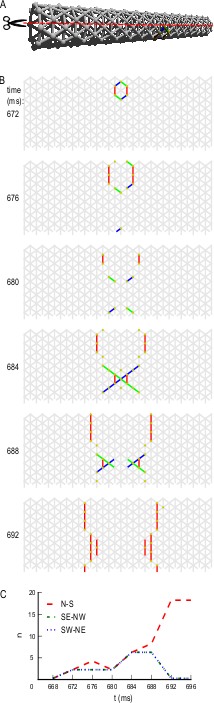}
%\end{center}
\caption{Wave patterns on a long tube (length: 32 cells, circumference: 8 cells, noise rate: 0.1 Hz). (A) Network geometry:  the scissors indicate the line at which the tube is cut for presentation in B. (B) Snapshots of network activity during 4 ms intervals in an illustrative phase of the dynamics. The wave fronts propagating transversely collide with each other, causing extinction due to the refractory period. Subsequently, the remaining wave-fronts propagating longitudinally dominate the dynamics and the North--South wave-front orientation dominates. (C) Temporal development of different wave-front orientations. Snapshot time markings are consistent with those in B.}
\label{fig:longpipe}
\end{figure}

In subpanel B of figure~\ref{fig:longpipe} we see the temporal development of wave-front patterns on the elongated cylindrical network also shown in 3D in subpanel A of the same figure. After initial excitation, the wave front grows uniformly in all directions until the wave fronts propagating transversely annihilate each other  opposite the wave front initiation point. What remains are two wave fronts propagating longitudinally. Provided no other noise induced wave fronts interfere with these wave fronts they will eventually reach the edge of the cylinder and disappear. In subpanel C of figure \ref{fig:longpipe}, where the wave-front orientation counts are shown for the time interval depicted in subpanel B, we clearly see the phenomena we just described reflected. Initially all wave-front counts grow at the same rate, then the South East--North West and South West--North East oriented wave fronts, that is, the wave fronts propagating transversely, annihilate each other. Subsequently, the North--South oriented wave fronts continue to grow in size until they become approximately stable during longitudinal propagation.

\begin{figure}
\begin{center}
\includegraphics[width=0.5\textwidth]{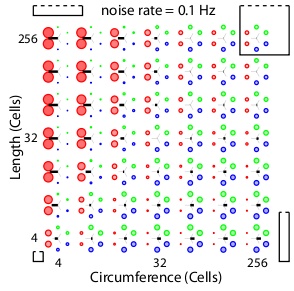}
\end{center}
\caption{Analysis of whole-body coordination for various body shapes. Relative wave-front orientation prevalences (represented by coloured disk diameter) and average propagation orientations (indicated with an oriented black bar) are shown for various body lengths and circumferences of the excitable myoepithelium. In the corners, the corresponding body shape is indicated with a rectangle. Bottom left: small `square' cylinder, top left: long cylinder with small circumference, top right: large `square' cylinder, bottom right: short cylinder with large circumference. This parameter scan shows three effects: (i) more elongated networks show better developed longitudinal wave fronts, (ii) whereas shorter networks show better developed transverse wave fronts, (iii) however for fixed length-to-circumference ratios (visible on the diagonals running from bottom left to top right) we can see that with increasing size preference for transverse or longitudinally moving wave fronts is lost.} 
\label{fig:WavefrontPrefsNoise0_1}
\end{figure}

To establish whether and under which conditions a surface of excitable cells generates coordinated patterns, we simulate our network at different network sizes and different noise rates. Our analysis methods allow us to screen a large parameter space for patterned activity. At an intermediate noise rate of 0.1 Hz figure~\ref{fig:WavefrontPrefsNoise0_1} shows the relative wave-front orientation prevalences (represented by the diameters of the coloured disks) and average propagation orientations (indicated by the orientation of the black bars) for various body lengths and circumferences of the excitable myoepithelium. In the corners of this figure we drew rectangles to illustrate the shape of the model network at the parameter settings used for the simulation in the corresponding corner. The ratio of circumference to length used in these drawings are understated with 1:1 (left bottom corner), 1:5 (left top corner), 5:5 (right top corner) and 5:1 (right bottom corner), while the actual ratios in the model network are 4:4, 4:256, 256:256 and 256:4, respectively. From the average propagation orientations in figure~\ref{fig:WavefrontPrefsNoise0_1} we see that there is a strong preference for longitudinal wave-front propagation if the model network axis is long compared to its circumference (upper left). In contrast, if the model network circumference is large compared to its axis (lower right), then we see a strong preference for transverse wave-front propagation. This is also visible from the relative wave-front orientation prevalences. Hence, we see that for these networks there is significant pattern formation. To extract how pattern formation scales with size we can study the change in pattern formation on the diagonals running parallel to the main diagonal (bottom-left corner to top-right corner). Along these diagonals body size varies while the ratio between axis length and circumference remains constant; the smallest body size is at the lower left side of these diagonals and the large body size at the upper right side. As we move along these diagonals to larger scale networks we observe that relative wave-front orientation prevalences equalize and average propagation orientation diminishes. This shows that pattern formation fails to reach network size when we move to large networks.

\begin{figure}
%\begin{center}
\includegraphics[height=0.8\textheight]{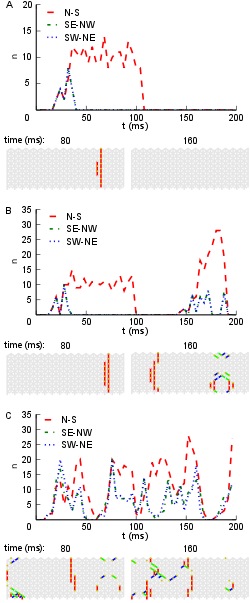}
%\end{center}
\caption{Development of wave-front orientation prevalences over time on a long tube (length: 32 cells, circumference: 8 cells) at three different noise rates. A. Low noise condition (0.01 Hz/cell). The graph at the top shows wave-front counts for all three orientations. It is clearly visible how activity first grows equally for all three orientations until the scale of activation pattern equals the tube's circumference, at which point in time the wave front moving longitudinally starts to dominate. B. At the intermediate noise rate (0.1 Hz/cell) we still observe growth of wave-front patterns to the scale of the animal, but occasionally several wave fronts are initiated in close succession leading to destructive interference. C. At a high noise rate (1 Hz/cell) wave fronts are initiated at a high rate and due to destructive interference with each other these wave fronts often fail to grow to the scale of the animal. As a result we no longer observe whole-body coordinated activity.} 
\label{fig:wavenoisetimelines}
\end{figure}

Noise, which represents spontaneous vesicle release, drives activity in our network, but we also expect it to interfere with emerging patterns. The reduction of the noise rate might therefore rescue patterning in large-scale networks, and an increase in noise rate might destroy patterning in small-scale networks. Figure~\ref{fig:wavenoisetimelines} illustrates the influence of the noise rate on whole-body coordination at a single fixed combination of length and circumference. In this figure each subpanel shows, for a specific noise rate, the temporal development of wave-front orientation prevalences. At the top of each subpanel we find the wave-front counts plotted versus time, followed by snapshots of the activity in the network. Subpanel A shows the low noise rate situation in which almost every excitation grows to network scale and induces coordinated activity. Subpanel B shows the intermediate noise rate situation in which many wave fronts grow to the short scale of the network but spreading on the long scale is often interrupted by collision with other wave fronts. Subpanel C shows the high noise-rate situation in which, in general, wave fronts are disrupted before reaching network scale and whole-body coordination at the network level is absent.

\begin{figure}
\begin{center}
\includegraphics[width=0.8\textwidth]{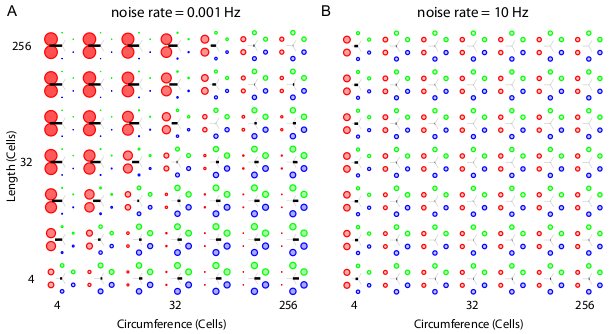}
\end{center}
\caption{Analysis of whole-body coordination for the extreme noise rates used in this study. Relative wave-front orientation prevalences (represented by the diameters of the coloured disks ) and average propagation orientations (indicated with the oriented of the black bar) are shown for various body lengths and circumferences of the excitable myoepithelium. The two subpanels are organized as in figure~\ref{fig:WavefrontPrefsNoise0_1}. (A) Low noise rate: 0.001 Hz, (B) High noise rate: 10 Hz. Compared to figure~\ref{fig:WavefrontPrefsNoise0_1} we find slightly stronger relative wave-front orientation prevalences and average propagation orientations at the low noise rate in panel (A). In contrast we clearly see the loss of whole-body coordination with increasing noise rates, as relative wave-front orientation prevalences become uniform and average propagation orientations are almost absent in panel (B).} 
\label{fig:WaveOrientation_NoiseRangeExtremes}
\end{figure}

Figure~\ref{fig:WaveOrientation_NoiseRangeExtremes} shows two parameter scans over network sizes, performed at the highest and the lowest noise rates used in this study. Both panels show relative wave-front orientation prevalences (represented by the diameters of the coloured disks) and average propagation orientations (indicated by the orientation of the black bars)  for various body lengths and circumferences of the excitable myoepithelium. The two subpanels are organized as in figure~\ref{fig:WavefrontPrefsNoise0_1} but the simulations were performed at a low noise rate of 0.001 Hz (A) and a high noise rate of 10 Hz (B). Compared to figure~\ref{fig:WavefrontPrefsNoise0_1}, we find slightly stronger relative wave-front orientation prevalences and average propagation orientations at the low noise rate in panel (A). In panel (B), in contrast, we see a loss of whole-body coordination with high noise rate, since relative wave-front orientation prevalences become uniform and average propagation orientations are almost absent at most parameter values in this scan. This shows that a reduction of chemical transmission noise can enhance the range of network size over which we find whole-body coordination, whereas an increase will reduce this range.  

\section*{Discussion}
To find answers to our main question, namely, how and why the first nervous systems evolved, we used a computational approach to investigate the possible behavioural relevance of basic forms of neural or proto-neural organization.

We argued first that two separate consecutive evolutionary steps introduced two key components of functional neuron anatomy:
\begin{enumerate}
\item[i.] Chemical transmission for intercellular electrical signalling to adjacent cells.
\item[ii.] Axodendritic processes that enable electrical signals to be sent to nonneighbouring cells.
\end{enumerate}
In addition, the origins of the first nervous systems are plausibly linked to the evolution of muscle, the key effector innovation that allowed animals to move even when they had grown to a large size~\cite{pantin1956origin,Seipel_Schmid_2005}.

The reasoning we used to construct our model consists of three main parts. First, we turned to modern myoepithelia as the most basic form of muscle coordination in existence today. There, coordination is brought about by an excitable epithelium which doubles as a contractile tissue, while the cells of this epithelium are directly connected by gap junctions. Second, we took into consideration that there is good evidence that gap junctions are a later evolutionary development compared to chemically induced action potentials. Third, we drew the conclusion that without electrical connections there would be a good evolutionary reason to develop chemical transmission as a first step towards the evolution of a basic proto-neural organization consisting of a myoepithelium relying on such chemical transmission.
 
Thus, our model allowed us to investigate the evolutionary possibility that such a proto-neural organization, which relies on chemical transmission but no axodendritic processes, played a role in organizing a basic form of muscle-based behaviour. We think that such a primitive form of muscle coordination is an essential intermediate stage preceding the next step towards full nervous systems, i.e. the development of axodendritic processes.

\subsection*{The emergence of whole-body coordination}
The simulation results showed that a proto-neural organization based on chemical transmission alone can lead to self-organized patterned activity at a whole-body scale, depending on the values of three main features: noise, body dimensions and body size. Let us briefly consider each of these features.

\emph{Noise:} Noise is required to initiate electrical activity. The cells included in our model epithelium showed spontaneous vesicle release that initiated action potentials which local chemical transmission subsequently spread out across the epithelium. This spontaneous activity could give rise to patterned activity at the body scale if the noise level was sufficiently low. At high noise levels, the subsequent spontaneous firing of action potentials disrupts already evolving large-scale patterns.

\emph{Body dimensions:} At sufficiently low spontaneous vesicle release rates the relative myoepithelium dimensions, i.e. the length to width ratio, determines the type of whole-body coordination. Wave fronts travelling along the short dimension die out either through collision with the edge or through annihilation with a wave front travelling in the opposite direction from the same initiation site. Wave fronts travelling along the long dimension then travel on and finally die out through the before-mentioned mechanism. At such vesicle release rates wave fronts travelling along the long dimension dominate the dynamics and lead to a primitive form of whole-body coordination. Thus body dimensions are a key feature for the emergence of body coordination under these circumstances.

\emph{Body size:} In our model the scale of these patterns is determined by the rate of spontaneous vesicle release. We see whole-body coordination emerge only when the scale of these patterns matches roughly with one of the dimensions of the animal, i.e. matches with length or circumference. Consequently, we see a reduction of whole-body coordination with the increase of animal size.

While the model remains rather basic, its significance lies in the way it shows the generic properties of a proto-neural myoepithelium without any additional features. The important message here is that such a system enables basic coordinated patterning on its own, without any sensory input, without any central pattern generators and without requiring any specific wiring or particular connections between the cells. Coordination can be cast as an ingrained self-organized feature of such an organization. At this stage there is no intrinsic need for particular neural `circuits' or wirings that require a specialized evolutionary route to explain their presence.

\subsection*{Evolutionary implications}
Our model shows how the first of the two evolutionary steps leading to full nervous systems could have resulted in an excitable myoepithelium that is roughly comparable to modern forms of excitable myoepithelia, but without direct electrical coupling. Given that current evidence suggests that such direct electrical coupling was a later evolutionary development, enabling early excitable epithelial systems can be cast as the primary evolutionary reason for the rise of the chemically induced transmission of electrical signals between neighbouring cells. Importantly, despite its similarity to modern non-neural myoepithelia, our model relies on local chemical transmission, which makes it a definite example of a proto-neural system. As such it is halfway between other excitable biological systems and modern nervous systems.

In addition, the generic capacity for patterning in this proto-neural organization remains limited to smaller body sizes and does not scale up to larger ones. If this limitation proves to be a genuine and general characteristic of this kind of organization, it will provide a major constraint and influence on our understanding of subsequent evolutionary developments such as the origins of full nervous systems with axodendritic processes and modern (myo)epithelia with gap junctions.

Given the present focus on the evolution of nervous systems the second option is perhaps less relevant. Still, our simulations can potentially  explain why electrical transmission became an important feature of modern excitable epithelia such as those in jellyfish: to reach whole-body or whole-organ coordination, a noise free coupling between the cells in these tissues is required, thus allowing the spreading of wave-fronts to the scale of the tissue.

More central to present purposes, the model contributes to our understanding of the evolution of early nervous systems and their precursors by suggesting new relevant details for their operation as well as a more gradual evolutionary path towards the modern neural organization. It will be interesting and necessary to look at additional features in follow-up research to investigate how the basic myoepithelial proto-neural organization can be evolutionarily improved upon to expand its coordinative features and reaction to stimuli. In this way, it also provides a good starting point for a further investigation of the other main step in nervous system evolution, i.e. axodendritic processes.

To conclude, the model suggests an evolutionary trajectory in which the first step of nervous system evolution provided an initial advantage for coordinating an extended contractile tissue in the absence of direct electrical connections between cells. At the same time, the model suggests that there are significant limitations to this initial organization, thus evolutionary pressure towards larger body size may have given rise to the next step in nervous system evolution. 

\subsection*{Follow-up questions}
The research described above suggest a number of general as well as specific questions and topics for further research. Here we will describe some of the most urgent and important follow-up questions. These will be divided into two methodological and two biological questions.

A first methodological issue concerns the problem of analyzing and measuring whole-body coordination as an emergent feature. So far, the primary metric has been the coherence of wave fronts with the width of a single cell. While this generally works for the sort of behaviour the current model exhibits, this is not a direct measure of activation relative to body size: it is merely a measure of wave-front coherence. To be able to say something about what activation does for the organism, it will be necessary to devise a measure that equates activity in areas relative to topology.

A second methodological issue concerns the need and possibility of adding movement and environmental dynamics to the basic model as presented here. Work on cardiomyocytes suggests that mechanics can be integrated into the model~\cite{NashHunter2000,nash2004electromechanical}. This would make it feasible to study interaction effects between wave propagation and changes in body shape. When combined with, for example, a fluid dynamics model for the environment this could provide insight into how early multicellulars may have moved in an aquatic environment.

From a biological perspective, the first important question concerns the various ways in which biological features can be added to a proto-neural myoepithelial organization. We will briefly mention a few possibilities of enriching such a myoepithelium:

\begin{itemize}
\item Central Pattern Generators (CPGs): Activity in the current model is initiated by random noise that sets individual cells firing and activity spreads from there. Patterns of activity derived from the interaction of such spontaneous wave fronts and the form of the animals are key factors for the global activity patterns. However, pacemaker cells or structures that initiate firing at a specific place and with a specific frequency might be able to entrain activity across the whole myoepithelium. It would be important to investigate how such CPGs can arise in a basic set up as described above, and how they impact on global myoepithelial activity.
\item Excitation and inhibition: Modern neurotransmitters can have either excitatory or inhibitory effects on the post-synaptic cell. The present model relies solely on excitation, but it would be interesting to see how adding different interaction dynamics can change the pattern dynamics.
\item Mechanosensory feedback: Modern cells contain mechanosensitive ion channels. These are ion channels which are activated by membrane stretch, and which in addition to several other functions, play a role in volume regulation and in the functioning of healthy muscle and cardiac cells~\cite{Hamill_2001}. Despite their general presence in modern cells their molecular identity is hard to establish electrophysiologically~\cite{Sachs_2010}, although recent developments show some progress~\cite{Gottlieb_2012,Coste_2012,Kim_2012}. Sufficient insight into the molecular identity of these mechanosensitive channels is a prerequisite for the kind of genetic analysis which is needed to hypothesize about their role in the evolution of neural systems. Including mechanosensitive channels in an excitable myoepithelium model which also includes the mechanics, could reveal their impact on whole-body coordination.
\item Sensory stimuli and feedback: Adding a way to initiate activity on the basis of external stimuli would also be an obvious way to extend the model. Note, however, that from the present perspective it is not sensory initiation but rather coordination of activity which is a prerequisite for whole body coordination and thus for  behaviour.
\end{itemize}

Another major biological question concerns the second step involved in the evolution of nervous systems namely the origin of elongated axodendritic processes. The proto-neural organization discussed so far produces limited coordinative abilities that can presumably be extended by some additional refinements. However, the evolution of processes provides the transition to modern nervous systems. A key question that must be answered is in what specific ways extended processes change the coordinative behaviour of such myoepithelia. While this question has so far proven to be extremely difficult to answer by more direct evolutionary approaches, the current modelling approach provides an excellent starting point for further investigation.

%\bibliography{scilib}

\newpage

% - A ------------------------------------------------------------------------------
\noindent
\begin{table}
\begin{tabularx}{\linewidth}{|l|X|}\hline
\hdr{2}{A}{Model Overview}\\\hline
\textbf{Cell populations}     & Single excitatory \\\hline
\textbf{Topology}         & Triangular lattice on a cylinder \\\hline
\textbf{Connectivity}     & Nearest-neighbour \\\hline
\textbf{Cell}     & Single-compartmental\\\hline
\textbf{Channels}   & Hodgkin-Huxley sodium and potassium\\\hline
\textbf{Chemical transmission}    & Double-exponential and conductance-based \\\hline
\textbf{Noise}    & Independent fixed-rate Poisson processes of spontaneous vesicle release\\\hline
\textbf{Measurements}     & Wave-front orientation \\\hline
\textbf{Simulator}        & NEURON (www.neuron.yale.edu) \\\hline
\end{tabularx}
\caption{Model Summary A: Overview}
\label{table:MS-A}
\end{table}
% - B -----------------------------------------------------------------------------
%
\noindent
\begin{table}
\begin{tabularx}{\linewidth}{|l|l|X|}\hline
\hdr{3}{B}{Cell and stimulus populations}\\\hline
  \textbf{Name}   & \textbf{Elements} & \textbf{Size} \\\hline
    HHCellList             &   HHCell   & $length*circumference$  \\\hline
	StimList &   Poisson process (NetStim)   & $length*circumference$  \\\hline
\end{tabularx}
\caption{Model Summary B: Cell and stimulus populations}
\label{table:MS-B}
\end{table}
% - C ------------------------------------------------------------------------------
%
\noindent
\begin{table}
\begin{tabularx}{\linewidth}{|l|l|l|X|}\hline
\hdr{4}{C}{Connectivity}\\\hline
\textbf{Name} & \textbf{Source} & \textbf{Target} & \textbf{Pattern} \\\hline
  NC\_HHCellList & HHCellList & HHCellList & 
  Nearest neighbour  \\\hline
  NC\_NS\_HHCellList & StimList  & HHCellList & 
  One-to-one  \\\hline
\end{tabularx}
\caption{Model Summary C: Connectivity}
\label{table:MS-C}
\end{table}
% - D ------------------------------------------------------------------------------
%
\noindent
\begin{table}
\begin{tabularx}{\linewidth}{|p{0.17\linewidth}|X|}\hline
\hdr{2}{D}{Cell Model}\\\hline
\textbf{Name} & HHCell \\\hline
\textbf{Type} & Single compartmental Hodgkin-Huxley model \\\hline
\textbf{Dynamics} &
\rule{1em}{0em}\vspace*{-3.5ex}
    \begin{eqnarray*}
      C_m \frac{dV_m}{dt}&=& -Ag_l(V_m-E_l)-A\bar{g}_K n^4(V_m-E_K) \\ 
						  &&- A\bar{g}_{Na} m^3 h (V_m-E_{Na})+I_{syn}\\
	\frac{dx}{dt}&=&-(x-x_\infty(V_m))/\tau_x \hspace{2 em} {\rm with \ } x=m,h,n\\
	\tau_x &=& 1/(\alpha_x + \beta_x) \hspace{3 em}   x_\infty = \alpha_x/(\alpha_x + \beta_x)
    \end{eqnarray*} 
\vspace*{-3.5ex}\rule{1em}{0em}
 \\\hline
\multirow{3}{*}{\textbf{Parameters}} & 
\rule{1em}{0em}\vspace*{-3.5ex}
\begin{eqnarray*}
%"m" sodium activation system and   "h" sodium inactivation system
		A&=& 400 \pi\ \mu m^2 = 1257\ \mu m^2\\
		C_m&=& A c_m = 12.57 pF\\
 		\bar{g}_{Na} &=& 0.12 {\rm \ S\  cm^{-2}},\hspace{1.3 em} E_{Na}= 50 {\rm \ mV}\\
		\alpha_m(V_m) &=&  \frac{-0.1(V_m+40)}{\exp(-(V_m+40)/10)-1)}\\
		\beta_m(V_m) &=&  4 \exp(-(V_m+65)/18)\\              
        \alpha_h(V_m) &=& .07  \exp(-(V_m+65)/20)\\ 
        \beta_h(V_m)&=& 1 / (\exp(-(V_m+35)/10) + 1)\\
%        :"n" potassium activation system
         \bar{g}_K &=& 0.036 {\rm \  S\  cm^{-2}},\hspace{2 em} E_K= -77 {\rm \ mV}\\
         \alpha_n(V_m) &=&  \frac{ -0.01 (V_m+55)}{\exp(-(V_m+55)/10)-1} \\ 
         \beta_n(V_m) &=& .125 \exp(-(V_m+65)/80)\\
%        :leak current
		  g_l &=& 0.0003 {\rm \  S\  cm^{-2}},\hspace{2 em}  E_l = -54.3 {\rm \ mV}
\end{eqnarray*} 
\vspace*{-3.5ex}\rule{1em}{0em}
\\ \hline
\end{tabularx}
\caption{Model Summary D: Single Cell Model}
\label{table:MS-D}
\end{table}
%
% - E------------------------------------------------------------------------------
\noindent
\begin{table}
\begin{tabularx}{\linewidth}{|l|X|}\hline
\hdr{2}{E}{Noise Model}\\\hline
\textbf{Type} & \textbf{Description} \\\hline
{Poisson process} & Fixed rate $\nu_{\text{spontaneous}}= 10^{3-noiseparameter} {\text \ Hz}$ 
 generator for each cell, the noise parameter is varied from $2$ to $6$.\\\hline
\end{tabularx}
\caption{Model Summary E: Noise Model}
\label{table:MS-E}
\end{table}
% - F ------------------------------------------------------------------------------
%
\noindent
\begin{table}
\begin{tabularx}{\linewidth}{|p{0.17\linewidth}|X|}\hline
\hdr{2}{F}{Chemical Transmission Model}\\\hline
\textbf{Name} & Exp2Syn \\\hline
\textbf{Type} & Double exponential conductance based \\\hline
\textbf{Dynamics} &
\rule{1em}{0em}\vspace*{-2.5ex}
    \begin{eqnarray*}
    I_{syn}&=&-(V_m-E_{syn}) \sum_{pre} w_{pre} \sum_{t_{pre}+d_{pre} \leq t} G(t-t_{pre}-d_{pre}) \\
     G(t) &=& G_0 \frac{\exp(-t/\tau_{decay})-\exp(-t/\tau_{rise})}{\exp(-t_{peak}/\tau_{decay})-\exp(-t_{peak}/\tau_{rise})}\\
     t_{peak}&=&\frac{\tau_{rise}\tau_{decay}}{\tau_{decay}-\tau_{rise}}
    \end{eqnarray*} 
\vspace*{-2.5ex}\rule{1em}{0em}
 \\\hline
\multirow{3}{*}{\textbf{Parameters}} & 
\rule{1em}{0em}\vspace*{-2.5ex}
\begin{eqnarray*}
G_0 &=& 1 \ \mu {\rm S}\\
w_{pre} &=& 0.001 {\rm \ or \ } 0\\ 
d_{pre} &=& 0.75 {\rm \ ms} \\
\tau_{rise} &=&  0.05 {\rm \ ms}\\
\tau_{decay} &=&  2 {\rm \ ms}\\
E_{syn} 	 &=& 0	{\rm \ mV} 
\end{eqnarray*} 
\vspace*{-2.5ex}\rule{1em}{0em}
\\ \hline
\end{tabularx}
\caption{Model Summary F: Chemical Transmission Model}
\label{table:MS-F}
\end{table}
%
% - G -----------------------------------------------------------------------------
\noindent
\begin{table}
\begin{tabularx}{\linewidth}{|X|}\hline
\hdr{1}{G}{Network Structure}\\\hline 
Triangular lattice on cylinder. The open ends are arbitrarily labelled East and West and both edges are aligned with the same primitive vector of the triangular lattice. Cell indices start at zero on a cell on the West edge. Indices are incremented by one for each lattice-constant-sized step on the edge in a direction arbitrarily marked as North. After labelling all the cells on this and subsequent rings, indexing continues stepping North from the first cell located North East of the last labelled cell until all cells are labelled. The number of cells on a single ring is specified by the circumference parameter, while the number of rings is specified by the length parameter.\\\hline
\end{tabularx}
\caption{Model Summary G: Network Structure}
\label{table:MS-G}
\end{table}
% - H -----------------------------------------------------------------------------
%
\noindent
\begin{table}
\begin{tabularx}{\linewidth}{|X|}\hline
\hdr{1}{H}{Analysis}\\\hline
Wave-front orientation preference is measured by counting how often two neighbouring cell pairs of a single orientation fire within $2$ ms of each other. In figure~\ref{fig:wavefrontanalysis} we develop several visual representations which capture this information. In addition, spike activities are shown as snapshots capturing $4$ ms of activity and as line graphs showing wave-front size as a function of time.  
\\\hline
\end{tabularx}
\caption{Model Summary H: Analysis}
\label{table:MS-H}
\end{table}

\renewcommand{\thefigure}{S. \arabic{figure}}
\setcounter{figure}{0} 
\begin{figure}[p]
\begin{center}
\includegraphics[height=0.85\textheight]{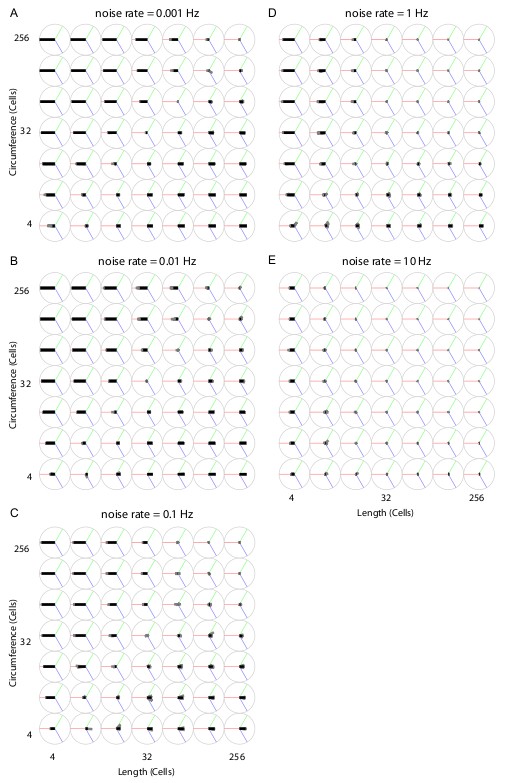}
\end{center}
\caption{Analysis of propagation direction orientation for various body shapes and noise rates. Propagation direction orientation, averaged over a unique combination of length, circumference and noise rate is indicated by an oriented black bar. There are 19 individual runs per unique combination of length, circumference and noise rate. Propagation direction of a single run is indicated with an oriented grey bar. These grey bars usually largely overlap with the black bar, indicating that these experiments are highly reproducible. The orientation calculation is explained in figure~\ref{fig:wavefrontanalysis}. (A) Noise rate: 0.001 Hz, (B) Noise rate: 0.01 Hz, (C) Noise rate: 0.1 Hz, (D) Noise rate: 1 Hz, (E) Noise rate: 10 Hz.} 
\label{fig:WavefrontIndividualRuns}
\end{figure}

\end{document}